\documentclass[manuscript]{acmart}

\usepackage{graphicx}
\usepackage{subcaption}
\usepackage{todonotes}
\usepackage{colortbl}
\usepackage{makecell}

\usepackage{amssymb}
\usepackage{pifont}        
\usepackage{enumitem}
\usepackage{soul}

\newcommand{\cmark}{\textcolor{green!60!black}{\ding{51}}}  
\newcommand{\xmark}{\textcolor{red}{\ding{55}}}   
\AtBeginDocument{%
  }

\setcopyright{acmlicensed}
\copyrightyear{2027}
\acmYear{2027}
\acmConference[CHI '27]{Proceedings of the 2027 CHI Conference on Human Factors in Computing Systems}{May 10--14, 2027}{Pittsburgh, PA, USA}
\acmBooktitle{Proceedings of the 2027 CHI Conference on Human Factors in Computing Systems (CHI '27), May 10--14, 2027, Pittsburgh, PA, USA}

\acmPrice{}
\newcommand{\squishlist}
{\begin{itemize}[itemsep=1pt,parsep=2pt,topsep=3pt,partopsep=0pt,leftmargin=0em, itemindent=1em,labelwidth=1em,labelsep=0.5em]}
\newcommand{\squishend}{\end{itemize}}
\newcommand{\squishenum}{\begin{enumerate}[itemsep=1pt,parsep=2pt,topsep=3pt,partopsep=0pt,leftmargin=0em, itemindent=1.5em,labelwidth=1em,labelsep=0.5em]}
\newcommand{\squishsubenum}{\begin{enumerate}[itemsep=1pt,parsep=2pt,topsep=0pt,partopsep=0pt,leftmargin=0em,listparindent=1.5em,labelwidth=1em,labelsep=0.5em]}
\newcommand{\squishenumend}{\end{enumerate}}

\begin{document}

\title{{\sysname}: Gaze-Anchored Spatial Perception Using Corneal Reflections}

\author{Seungjoo Lee}
\affiliation{%
  \institution{Carnegie Mellon University}
  \city{Pittsburgh, PA}
  \country{USA}}
\email{seungjoolee@cmu.edu}

\author{Vimal Mollyn}
\affiliation{%
  \institution{Carnegie Mellon University}
  \city{Pittsburgh, PA}
  \country{USA}}
\email{vmollyn@cs.cmu.edu}

\author{Chris Harrison}
\affiliation{%
  \institution{Carnegie Mellon University}
  \city{Pittsburgh, PA}
  \country{USA}}
\email{chris.harrison@cs.cmu.edu}

\author{Justin Chan}
\affiliation{%
  \institution{Carnegie Mellon University}
  \city{Pittsburgh, PA}
  \country{USA}}
\email{justinchan@cmu.edu}

\author{Mayank Goel}
\affiliation{%
  \institution{Carnegie Mellon University}
  \city{Pittsburgh, PA}
  \country{USA}}
\email{mayankgoel@cmu.edu}

\renewcommand{\shortauthors}{Lee et al.}

\newcommand{\sysname}{GlintMarkers}

\begin{abstract}
AI assistants on smart glasses need to know what surrounds the user and what the user is looking at. Obtaining this context typically relies on a world-facing camera, raising privacy concerns for bystanders. We present {\sysname}, a system for gaze-anchored spatial perception using a single inward-facing eye camera. Our key observation is that the cornea acts as a mirror that encodes both gaze direction and visual information about the environment in a small, low-contrast reflection. To extract spatial and semantic information from this reflection despite the camera's limited pixel budget, we design passive retroreflective markers that concentrate reflected near-infrared light onto the cornea, producing bright glint patterns. We develop a custom Perspective-n-Point (PnP) estimation framework adapted to corneal imaging and perform orientation and distance estimation of marked objects, as well as unique object identification.
\end{abstract}

\ccsdesc[500]{Human-centered computing~Ubiquitous and mobile computing systems and tools}

\keywords{Corneal reflections, smart glasses, gaze interaction, extended reality, augmented reality}

\begin{teaserfigure}
\centering
  \includegraphics[width=\textwidth]{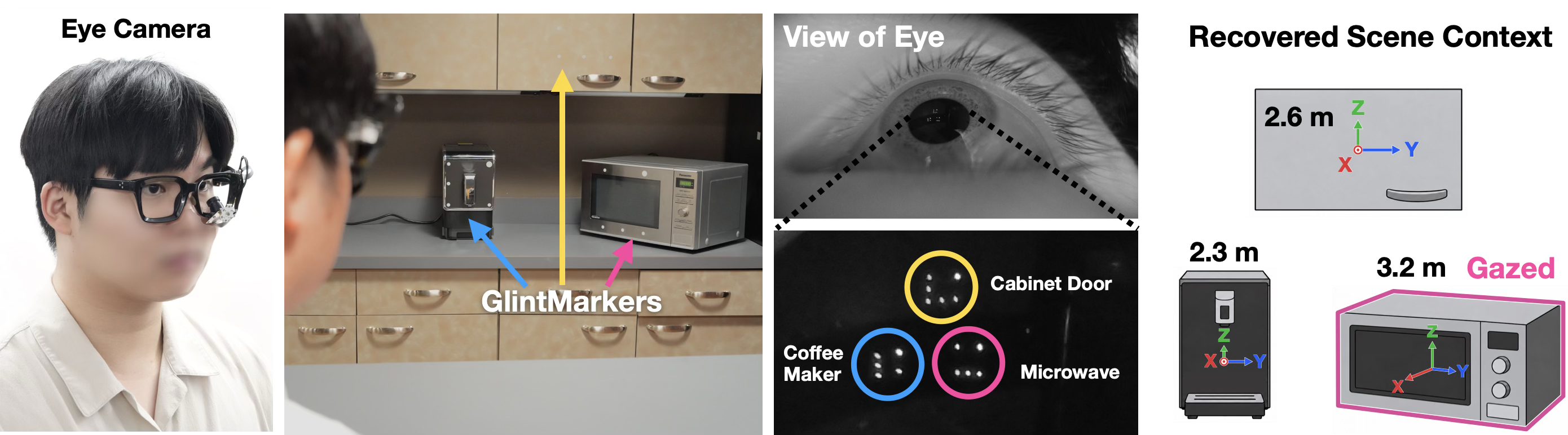}
  \vspace{-1.7em}
  \caption{{\bf {\sysname} enables spatial perception of the physical world using only an inward-facing eye camera. By capturing corneal reflections of {\sysname} placed on everyday objects, the system identifies marked objects, estimates their 3D orientation and distance relative to the user, and determines which one the user is gazing at.}}
  \vspace{1em}
  \label{fig:teaser}
\end{teaserfigure}

\maketitle
\section{Introduction}

Knowing what is in the scene and what the user is looking at~\cite{konrad2024gazegpt, wilson2025gaze} enables AI assistants on smart glasses~\cite{raybanmeta_lookandask, satori2025, sensibleagent2025} to provide context-aware assistance.
To sense the scene, two approaches are widely used: recognizing objects by appearance with computer vision~\cite{promemassist2025, satori2025}, for example an object detector such as YOLO~\cite{redmon2016you}, or placing fiducial markers such as QR codes or ArUco markers on objects~\cite{iso18004, garrido2014aruco, olson2011apriltag, 803809}.
Both approaches typically rely on a world-facing camera, raising significant privacy concerns~\cite{privaceye2019, denning2014bystanders, koelle2018bothers}.
In a survey, 65--90\% of bystanders reported they would take defensive action against glasses with cameras across the scenarios studied~\cite{wang2026mindgap}.

In this work, we explore an alternative approach that recognizes which objects are in a scene, estimates their orientation and distance, and identifies which one the user is looking at, all without a world-facing camera.
We build on eye-tracking cameras already integrated into head-worn devices~\cite{eyesurvey}: HoloLens~2 and Apple Vision Pro use gaze for input~\cite{hololens2_eyetracking, visionpro}, and glasses prototypes such as Meta's Orion and Aria Gen~2 include eye cameras~\cite{orion, ariagen2_spec}.
The images these cameras capture carry two signals simultaneously: the pupil position supports gaze estimation~\cite{kassner2014pupil}, and because the cornea acts as a convex mirror, the same image contains a small reflection of the scene in front of the user~\cite{nishino2004eyes, nishino2006corneal} (Fig.~\ref{fig:teaser}).

We present \textit{\sysname}, a system that enables gaze-anchored spatial perception from this reflection.
Our passive markers, called {\sysname}, consist of small retroreflective patches designed to be decoded from corneal reflections.
A near-infrared (NIR) LED on the glasses illuminates the markers.
Retroreflective material concentrates the reflected light back toward its source rather than spreading it in all directions, so each patch produces a bright point reflection that appears as a sharp \textit{glint} in the corneal reflection.
From these glints, {\sysname} recovers the identity, three-degree-of-freedom (3-DoF) orientation, and distance of each marked object in real time, and it determines which object the user is gazing at from the pupil position.
To our knowledge, {\sysname} is the first system to recover these quantities from passive markers via the corneal reflection captured by an eye-facing camera.

The key technical challenges are that the entire scene is compressed onto the small surface of the cornea, which has a reported reflectivity below 1\%~\cite{nishino2004eyes} and is captured with a limited pixel budget~\cite{nitschke2013corneal}, and that the camera observes a virtual image formed by a curved mirror rather than the scene itself.
To address these challenges, we make three technical contributions:
\squishlist
\item We introduce a passive retroreflective marker for corneal sensing -- a sparse grid of retroreflective patches that encodes an object identifier and serves as a fiducial for orientation and distance estimation.
\item We build a proof-of-concept eyewear prototype and a real-time pipeline that detects marker glints in the eye image, estimates 3-DoF orientation and distance with Perspective-n-Point (PnP) estimation adapted to the cornea's virtual image, and identifies the attended marker from the pupil position after a one-time per-user calibration.
\item We evaluate the system with ten participants. The system identifies the gazed object among six visible {\sysname} with 94--99\% per-frame accuracy, estimates distance with a mean absolute error of 0.120\,m over 2--4\,m, and recovers roll, yaw, and pitch with errors of approximately 3$^\circ$, 8--10$^\circ$, and 9--10$^\circ$, respectively. With six additional participants, orientation and distance remained recoverable for {\sysname} 10$^\circ$ and 20$^\circ$ away from the point of gaze (distance error 0.143\,m).
\squishend


\definecolor{Gray}{gray}{0.95}
\begin{table*}[t]
\centering
\resizebox{\textwidth}{!}{%
\begin{tabular}{l c c l c c c c}
\toprule
\textbf{\begin{tabular}[l]{@{}l@{}}System \\ Name\end{tabular}} &
\textbf{\begin{tabular}[c]{@{}c@{}}No World-Facing \\ Camera\end{tabular}} &
\textbf{\begin{tabular}[c]{@{}c@{}}Passive \\ Marker\end{tabular}} &
\textbf{\begin{tabular}[l]{@{}l@{}}Marker \\ Properties\end{tabular}} &
\textbf{\begin{tabular}[c]{@{}c@{}}Estimates \\ Identity\end{tabular}} &
\textbf{\begin{tabular}[c]{@{}c@{}}Estimates \\ Orientation\end{tabular}} &
\textbf{\begin{tabular}[c]{@{}c@{}}Estimates \\ Distance\end{tabular}} &
\textbf{\begin{tabular}[c]{@{}c@{}}Reported \\ Range\end{tabular}} \\
\midrule
RetroSphere~\cite{10.1145/3569479} & \xmark & \cmark & Retroreflective spheres & \xmark & \cmark & \cmark & 1\,m \\
InfraredTags~\cite{10.1145/3491102.3501951} & \xmark & \cmark & Code under IR-passing shell & \cmark & \cmark$^{\dagger}$ & --- & 2.5\,m \\
BrightMarker~\cite{10.1145/3586183.3606758} & \xmark & \cmark & Fluorescent 3D print & \cmark & \cmark & \cmark & $>2$\,m (tracking) \\
Koshikawa et al.~\cite{koshikawa2022transparent} & \cmark & \cmark$^{*}$ & Half-wave film on LCD & \cmark & \xmark & \xmark & 0.7\,m (LCD) \\
ViewPointer~\cite{viewpointer2005} & \cmark & \xmark & Powered IR LEDs & \cmark & \xmark & \xmark & Up to 3\,m \\
\rowcolor{Gray}
{\sysname} (ours) & \cmark & \cmark & Retroreflective patches & \cmark & \cmark & \cmark & Up to 4\,m \\
\bottomrule
\end{tabular}}
\par\vspace{1mm}
\begin{minipage}{\textwidth}
\footnotesize\raggedright
--- = not reported; ranges are study-specific.
$^{*}$Unpowered film requiring an illuminated LCD.
$^{\dagger}$Wheel orientation demonstrated; full 3-DoF orientation not reported.
\end{minipage}
\vspace*{2mm}\caption{Comparison of selected marker-based identification and tracking systems. {\sysname} is the only system in this comparison that uses passive markers and an eye-facing camera to recover object identity, 3-DoF orientation, and distance.}
\label{tab:rw_compare}
\end{table*}

\section{Related Work}

{\sysname} connects object identification and spatial tracking with gaze sensing through an eye-facing camera. We situate this combination within work on visual context sensing, passive and active tags, and corneal imaging. Table~\ref{tab:rw_compare} compares selected marker-based systems discussed below.

\subsection{Vision-Based Context and Gaze Sensing}
\label{sec:rw_vision}

Vision-based systems use world-facing cameras to sense objects, activities, and ongoing tasks. This context supports interactive digital content linked to physical objects~\cite{dogan2024xrobjects}, inference of user goals~\cite{satori2025}, and proactive or step-specific assistance~\cite{sensibleagent2025, promemassist2025, pro2assist2026}.

Adding gaze helps determine which objects the user is attending to or referring to. Combining gaze with scene imagery supports object recognition~\cite{ishiguro2010aided, toyama2012gaze}, query interpretation~\cite{konrad2024gazegpt, wang2024gvoila, rekimoto2025gazellm}, and disambiguation of references to physical objects~\cite{gazepointar2024}.

However, continuous visual sensing incurs power and processing costs on wearable devices~\cite{likamwa2014glass, proagent2025}. Recognition based on appearance can also struggle to distinguish visually identical object instances~\cite{blearvis2023}.

World-facing cameras also raise bystander privacy concerns, including discomfort about recording and identification, uncertainty about camera use, and a preference for being asked beforehand~\cite{denning2014bystanders}. In a recent survey, most bystanders reported that they would take defensive action, such as avoiding the camera or asking the wearer to stop recording~\cite{wang2026mindgap}.

These limitations motivate {\sysname}'s design: recover object identity and spatial context through an eye-facing camera while retaining gaze as an input for interaction. Passive markers make this possible without a world-facing camera.

\subsection{Passive Markers for Identification and Tracking}
\label{sec:rw_markers}

Passive markers link objects to digital information without relying on object appearance~\cite{want1999tags}. QR codes~\cite{iso18004} encode data, while fiducial systems such as ARToolKit~\cite{803809}, AprilTag~\cite{olson2011apriltag}, and ArUco~\cite{garrido2014aruco} support identity and pose estimation. Dot-based markers and retroreflective constellations likewise support identification or geometric tracking~\cite{bergamasco2011runetag, bergamasco2013pitag, pintaric2008rigid}. RetroSphere~\cite{10.1145/3569479}, for example, tracks controllers in 6-DoF using retroreflective spheres and headset-mounted infrared cameras.

Passive optical patterns also reveal motion: overlapping printed patterns amplify small widget displacements for camera readout~\cite{camposzamora2024moirewidgets}, retroreflector speckle supports rotation and depth estimation with a lensless sensor~\cite{chen2024spectrack}, and laser measurements of sticker vibrations reveal appliance operation and activities~\cite{vibrosight2018}.

Other markers limit the visibility of codes while preserving readout by world-facing NIR cameras. They use infrared-transmissive coverings~\cite{10.1145/3491102.3501951}, fluorescent patterns within printed objects~\cite{10.1145/3586183.3606758}, or infrared inkjet watermarks on printed documents~\cite{feick2025imprinto}.

To localize objects hidden from view, X-AR~\cite{boroushaki2023xar} augments visual tracking with an RFID reader that detects passive tags.

The optical systems above observe markers directly using world-facing cameras or dedicated readers. {\sysname} also uses passive markers, but its sparse retroreflective layout produces distinguishable corneal glints, enabling spatial sensing through an eye-facing camera already used for eye tracking.

\subsection{Active Tags for Spatial Interaction}
\label{sec:rw_active}

Active tags help distinguish objects in world-facing camera images. ID CAM~\cite{idcam2003} identifies objects from blinking LED beacons using a high-speed camera, while BLEARVIS~\cite{blearvis2023} combines Bluetooth direction finding with visual recognition to distinguish visually identical objects.

Active tags also communicate device and sensor information. LightAnchors~\cite{lightanchors2019} and InfoLED~\cite{infoled2019} track LEDs in world-facing images to anchor AR content while their blinking signals transmit identifiers and device information. LuxAct~\cite{Luxact} powers LEDs through user interactions, conveying identity and sensor data to an RGB camera without batteries.

HOBS~\cite{hobs2014} and InSight~\cite{insight2017} let users select objects by turning their heads toward them. Both use an infrared emitter on the glasses and powered receivers on objects, without requiring a world-facing camera. EchoSight~\cite{echosight2025} further supports identification and two-way communication through an infrared reader on the glasses and electronically modulated retroreflective tags.

These systems require electronics at the object for identification, communication, or localization. {\sysname} uses a passive layout for identity and spatial sensing, with an illuminator and eye camera on the glasses.

\subsection{Sensing Through Corneal Reflections}
\label{sec:rw_corneal}

Corneal reflections capture surrounding scene information in a small image region, where low contrast and iris texture can make scene details difficult to distinguish~\cite{nitschke2013corneal, wang2008separating}. Prior work has estimated surrounding lighting conditions~\cite{nishino2004eyes} and display position and orientation~\cite{nitschke2011display, schnieders2010reconstruction}, enhanced reflected image resolution~\cite{nitschke2012super}, reconstructed 3D scenes from color portrait sequences~\cite{alzayer2024seeing}, and recognized bystanders~\cite{jenkins2013identifiable}. Recognizable scene information has been recovered from visible-light reflections under specific capture conditions, such as high-resolution color photography, flash illumination, or short eye-to-target distances (approximately 1~m in the bystander study).

Corneal images also support gaze estimation~\cite{nitschke2013isee, lander2017eyemirror} and attended-object recognition~\cite{takemura2014estimation}. HiFiGaze~\cite{hifigaze2026} uses known screen content to locate its reflection and improve screen-relative gaze estimation. Wearable RGB eye recordings have supported corneal lifelogging~\cite{lifelogging1} and post-hoc recognition of attended and peripheral objects~\cite{lander2017attention}.

Other systems estimate gaze from reflections of projected infrared patterns~\cite{nakazawa2012point}, active LED markers~\cite{nakazawa2017corneal}, or polarization-encoded screen markers~\cite{koshikawa2022transparent}. ViewPointer~\cite{viewpointer2005} reads battery-powered infrared tags through a filtered eye camera to identify the gazed object.

{\sysname} builds on corneal marker sensing with passive retroreflective layouts illuminated from the glasses. Its NIR-filtered eye camera follows common eye-tracking designs~\cite{kassner2014pupil, lee2015gaze}, suppressing visible-light scene reflections while preserving marker glints. Relative glint positions provide object identity, 3-DoF orientation, and distance alongside gaze-based selection.

\begin{figure}[t]
    \centering
    \includegraphics[width=0.4\linewidth]{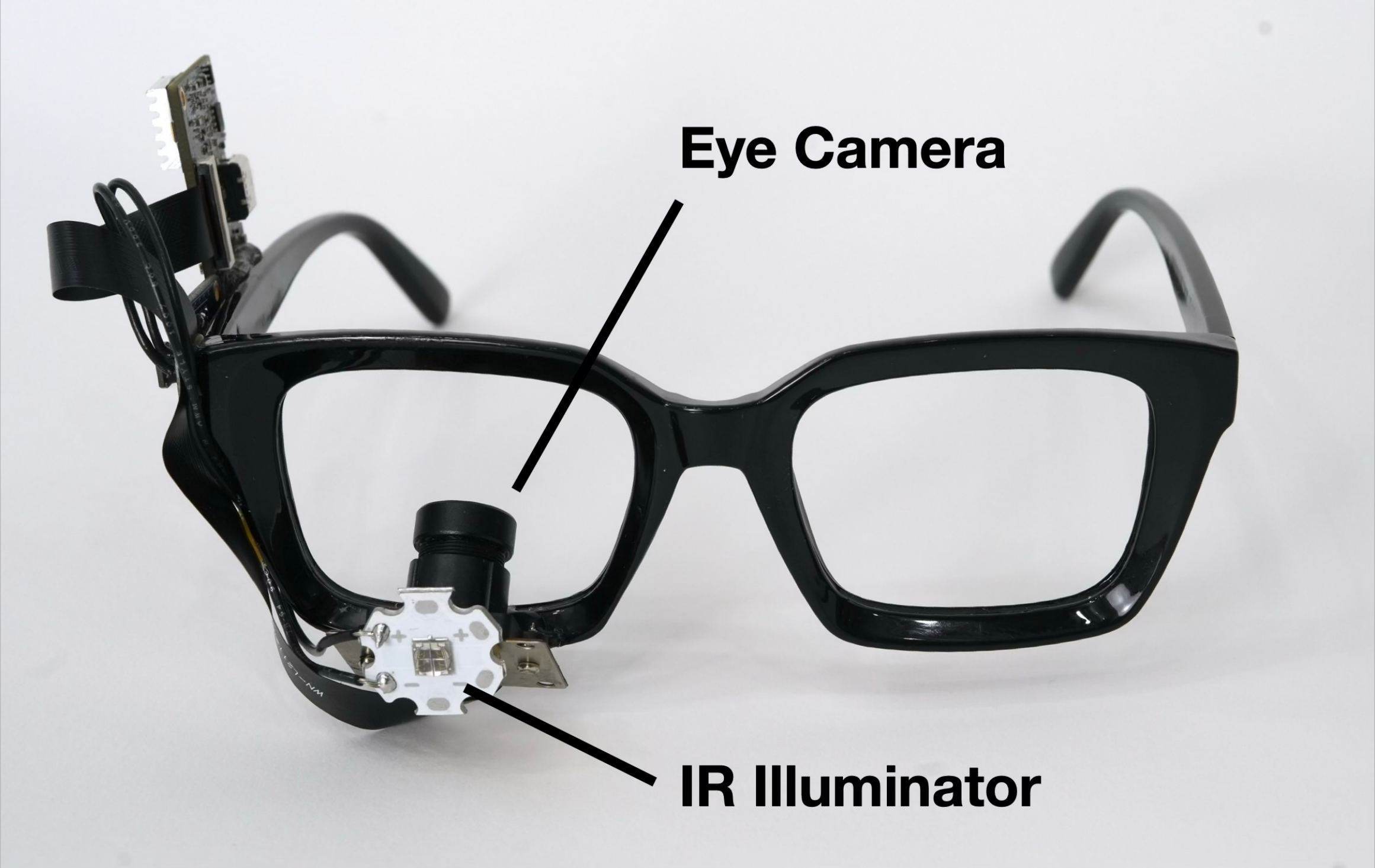}
    \caption{{\sysname} glasses prototype.}
    \label{fig:hardware}
\end{figure}

\section{Glasses Prototype}

We built a proof-of-concept glasses prototype with an eye-facing camera and a near-infrared (NIR) illuminator directed toward the scene (Fig.~\ref{fig:hardware}). 
To capture corneal reflections, we mounted an RGB camera~\cite{aliexpress_eye_camera} (3840 $\times$ 2160 resolution, 15 FPS) on the lower rim of a commodity eyeglass frame, oriented upward toward the eye. We replaced the camera's built-in IR-cut filter with a MidOpt BP850 NIR bandpass filter (manufacturer-specified useful range: 820--910~nm)~\cite{midopt_bp850} to suppress visible-light reflections.

The prototype uses a 1~W, 850~nm NIR LED with a $120^{\circ}$ viewing angle~\cite{amazon_ledguhon_850nm_led}. To drive the LED, we used a constant-current LED driver board~\cite{ak-led-driver}. The camera data were streamed over USB to a nearby laptop (MacBook Pro with an M4 Pro) for processing.

\section{Design of {\sysname}}
\begin{figure*}[t]
    \centering
    \includegraphics[width=0.9\linewidth]{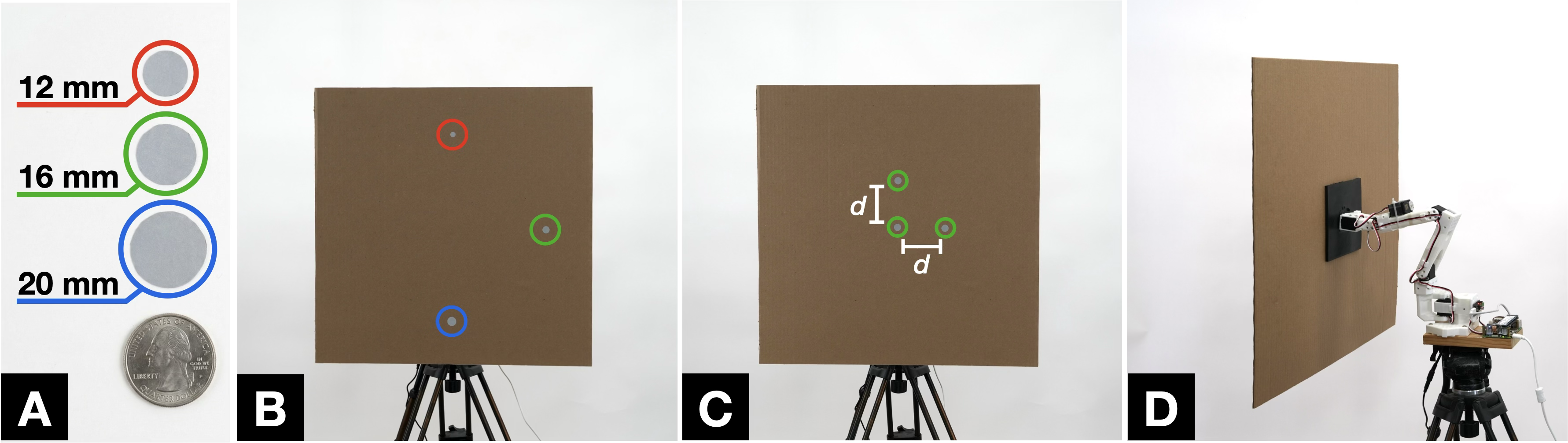}
    \caption{Experimental setup for characterizing retroreflective marker design parameters. (A) Retroreflective patches of three diameters (12, 16, and 20~mm). (B) Setup for evaluating patch size. (C) Setup for evaluating inter-patch spacing $d$ using an L-shaped arrangement. (D) The marker board is mounted on a robotic arm that sweeps through a range of orientations while participants fixate on the board at different distances.}
    \label{fig:patches}
\end{figure*}
{\sysname} are composed of multiple retroreflective patches which are designed to be visible within corneal reflections. Designing an effective marker requires balancing two competing concerns: individual patches must be large enough to appear as distinct, separable reflections in the corneal reflection, but the overall marker must remain small enough to attach to everyday objects.
To characterize this tradeoff, we built an automated data collection platform (Fig.~\ref{fig:patches}) in which a marker board was mounted on a robot arm that swept through a range of orientations: pitch within $\pm~30^{\circ}$, yaw within $\pm~45^{\circ}$, and roll within $\pm~45^{\circ}$, at approximately $20^{\circ}$/s. We conducted the characterization studies under normal indoor lighting.

\subsection{Effect of Retroreflective Patch Size} 
To characterize the effect of patch size, we conducted a preliminary study with five participants following human-subjects approval. We placed three retroreflective patches of different sizes (12, 16, and 20~mm) on a board, with sufficient spacing between them to avoid interference. The participants were asked to keep their head directed at the center of the board and fixate on it while the robot arm swept through various poses at distances of 2, 3, 4, and 5~m. Each sweep lasted 1~minute, yielding 900 eye images per distance.

\begin{figure}[t]
    \centering
    \begin{subfigure}[t]{0.25\linewidth}
        \centering
        \includegraphics[width=\linewidth]{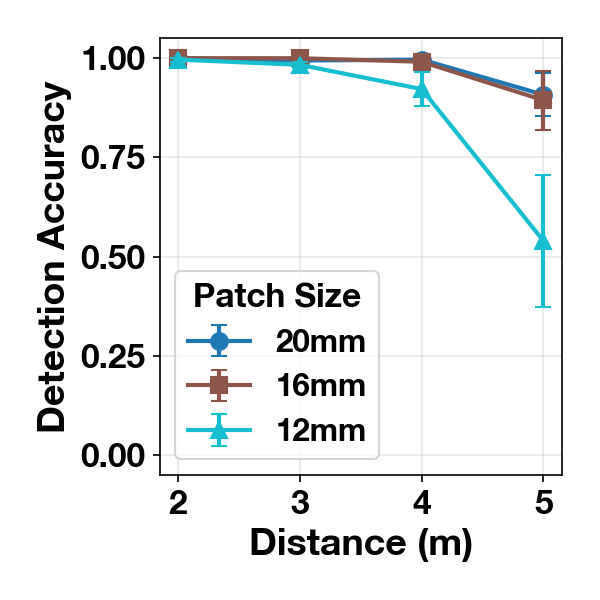}
    \end{subfigure}
    \begin{subfigure}[t]{0.25\linewidth}
        \centering
        \includegraphics[width=\linewidth]{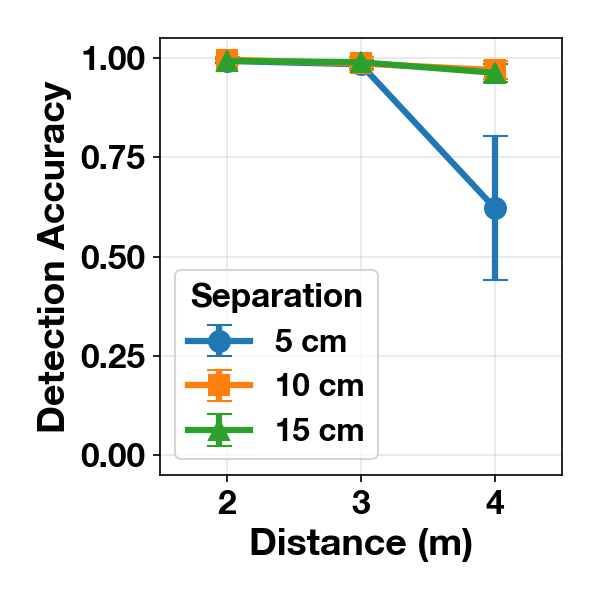}
    \end{subfigure}
    \vspace{-3mm}
    \caption{Left: Detection accuracy of different sized retroreflective patches over distances of 2 - 5~m. Right: Detection accuracy of different spaced retroreflective patches (16~mm size) over distances of 2 - 4~m. 
    }
    \label{fig:eval_size_spacing}
\end{figure}

We assess the detection accuracy across three patch sizes, where a marker is considered correctly detected if each individual patches are visible in the corneal reflections to a human annotator. Frames affected by blinks were excluded from analysis.
Fig.~\ref{fig:eval_size_spacing} shows that the 16 and 20~mm patch had high detection accuracies at 2 to 4~m from 99.1 to 100\%, decreasing to 89.4 to 90.8\% at 5~m. The detection accuracy of the 12~mm patch degraded sooner, with accuracy falling to 92.2\% at 4~m.

These results show that all three patch sizes are suitable for interaction within 4~m. Since patch size has little effect on the overall marker footprint, we select 16~mm for the longer-range marker design.

\begin{figure}[t]
    \centering
    \includegraphics[width=0.6\linewidth]{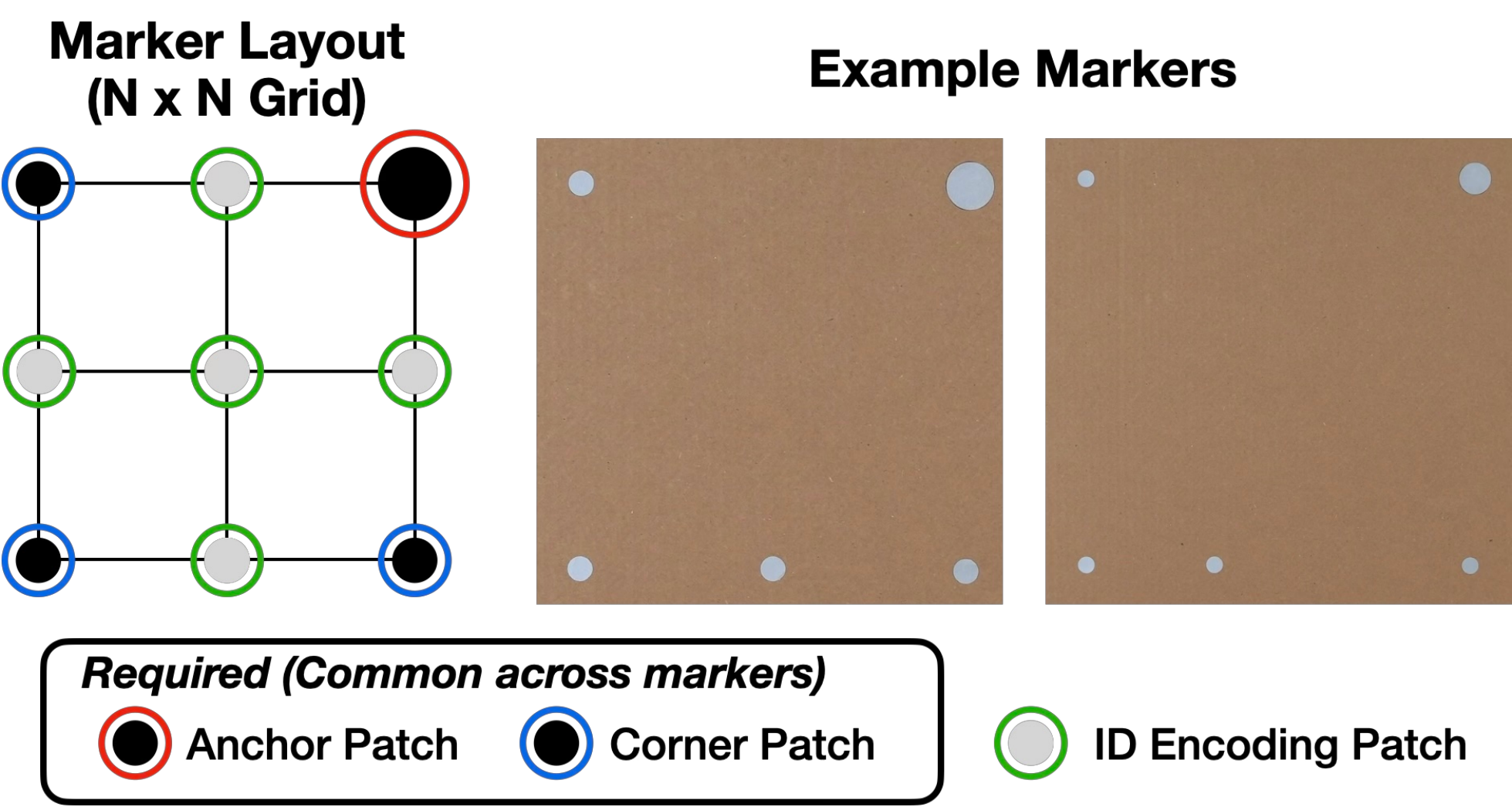}
    \caption{{\sysname} layout. An $N\times N$ grid of retroreflective patches encodes object ID and provides fiducial anchors for PnP pose estimation (3-DoF orientation and distance). Example $3 \times 3$ and $4 \times 4$ layouts are shown.}
    \label{fig:marker_layout}
\end{figure}

\subsection{Effect of Retroreflective Patch Spacing} 
Next, we conducted a second study with the same five participants to characterize the effect of patch spacing on detection accuracy. We evaluated three patch spacings (5, 10, and 15~cm) using 16~mm patches. For each spacing, we arranged three patches in an L shape, with equal horizontal and vertical spacing. Similar to the previous study, participants were asked to fixate on the board while it was placed at distances ranging from 2 to 4~m.

We scored each frame using a Difference-of-Gaussians (DoG) blob detector and considered a marker detectable when all three blobs were detected and their pairwise distances were within 60~px.

Our results in Fig.~\ref{fig:eval_size_spacing} show that detection accuracy remained consistently high for 10 and 15~cm patch spacings, ranging from 96.3\% to 99.6\% across distances of 2--4~m. In contrast, accuracy decreased at 4~m when the patch spacing was reduced to 5~cm.

\subsection{Marker Layout Design}

Based on these findings, we design two {\sysname} variants: a short- to mid-range marker using 12~mm patches with 5~cm spacing (2--3~m), and a longer-range marker using 16~mm patches with 10~cm spacing (up to 4~m).
Both variants use an $N \times N$ grid (Fig.~\ref{fig:marker_layout}), and we evaluate $3 \times 3$ and $4 \times 4$ configurations.

We place retroreflective patches at all four corners to serve as fiducial anchors for PnP-based pose estimation~(Section~\ref{section:id_and_pose}). Matching these four patches to their glints defines the planar image transform required by our pose estimator~\cite{collins2014infinitesimal}. To resolve rotational ambiguity, the top-right corner patch is larger than the others, establishing a canonical orientation. We use a 20~mm orientation patch for the short- to mid-range marker and a 30~mm orientation patch for the longer-range markers.

With the corners reserved for pose estimation, the remaining interior grid positions are available for identity encoding. The presence or absence of a retroreflective patch at each position acts as a binary bit, yielding 32 possible binary configurations for $3 \times 3$ configuration. The same design principle can be extended to larger grids to support a greater number of uniquely identifiable objects.

We use a grid for simplicity, although other patch layouts on a flat surface could also support object identification and pose estimation.

\begin{figure*}[t]
    \centering
    \includegraphics[width=0.8\linewidth]{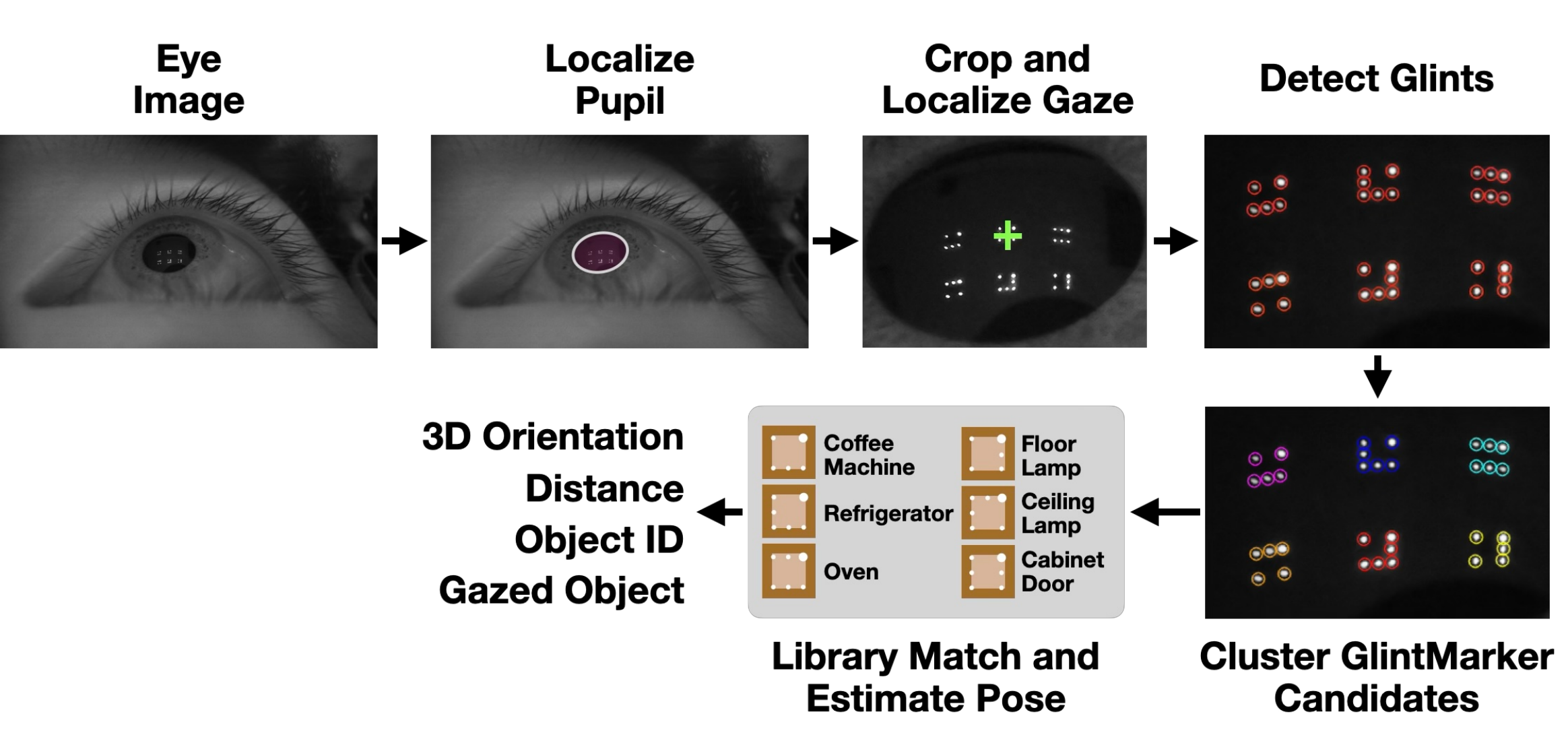}
    \vspace{-1em}
    \caption{High-level overview of {\sysname}'s processing pipeline.}
    \label{fig:pipeline}
\end{figure*}

\section{Software Implementation}
\label{sec:software}

{\sysname} localizes the pupil and gaze coordinate, then detects and clusters glints. It matches each cluster against a library of registered objects to identify the object and estimate its orientation and distance (Fig.~\ref{fig:pipeline}).

\subsection{Pupil and Gaze Localization}

We correct lens distortion using the eye camera's calibrated intrinsics. Our prototype uses DeepVOG~\cite{YIU2019108307} to produce a pupil probability map, which we threshold before fitting an ellipse to its largest connected component. Low-confidence detections are flagged as blinks. The ellipse defines the region for glint detection. Other pupil detectors could provide the center and boundary required by the subsequent stages.

Adding a calibrated user offset to the pupil center gives a 2D gaze coordinate in the eye image (Sec.~\ref{sec:calibration}). The crosshair in Fig.~\ref{fig:pipeline} marks this coordinate.

\subsection{Glint Detection and Clustering}

We restrict glint detection to the pupil region, which provides a dark background without the surrounding iris texture.

A Difference-of-Gaussians blob detector~\cite{blob} localizes candidate glints. Because the reflected patterns occupy a small image region, small localization errors can affect the estimated orientation and distance. We therefore refine each glint center to sub-pixel precision using an intensity-weighted centroid after subtracting the local background. Each refinement window is clipped to the glint's Voronoi region, excluding pixels closer to a neighboring glint and reducing interference between nearby reflections. We then group the refined centers using DBSCAN~\cite{schubert2017dbscan}. Clusters containing at least four glints become candidate markers.

\subsection{Marker Identification and Pose Estimation}
\label{section:id_and_pose}

The library stores each object's ID and its marker's patch layout and spacing. We fit these layouts to each glint cluster to identify the object and estimate the marker's pose.

To fit a marker layout, we first determine which patch produced each detected glint. The four glints farthest from the cluster centroid are candidate corners. We treat the brightest corner as the larger anchor patch and order the corners clockwise from it. We use the matched corners to compute an image transform (homography) that predicts where the other patches should appear.

Using the matched patch and glint positions, Perspective-n-Point (PnP) estimates the marker's pose as a rotation matrix $\mathbf{R}$ and a translation vector $\mathbf{t}$. We use Infinitesimal Plane-based Pose Estimation (IPPE)~\cite{collins2014infinitesimal} because it efficiently estimates pose from coplanar points. Reprojection RMSE is the root-mean-square pixel distance between detected glint centers and their positions predicted by the estimated pose.

We prioritize layouts whose patch count matches the detected glint count and use subset matching to account for spurious glints from surface reflections. Most clusters contain no spurious glints; when present, there are usually only one or two, keeping the subset search small. A candidate match is accepted only if its reprojection RMSE is below a threshold. If multiple candidates pass, we select the one with the lowest error.

IPPE can return two poses with similar reprojection errors. We consider both reprojection error and rotation changes between consecutive frames to favor poses consistent with previous estimates and reduce switching.

After identifying a marker, we track its glints across frames using pyramidal Lucas--Kanade optical flow~\cite{lucas1981iterative} and update its pose from the tracked positions. If tracking is lost during a blink or rapid head movement, we restart from glint detection.

The curved cornea forms a virtual image, so PnP approximates the reflected geometry. We calibrate the rotation estimate $\mathbf{R}$ to obtain 3-DoF orientation and convert the translation magnitude $\|\mathbf{t}\|$ to marker-to-user distance (Sec.~\ref{sec:calibration}). The individual translation components do not provide a stable world-space object position.

The cluster whose centroid is nearest the gaze coordinate supplies the gazed-object identifier. Identity, orientation, and distance are estimated for every accepted marker, regardless of gaze.

\begin{figure*}[t]
    \centering
    \includegraphics[width=0.8\linewidth]{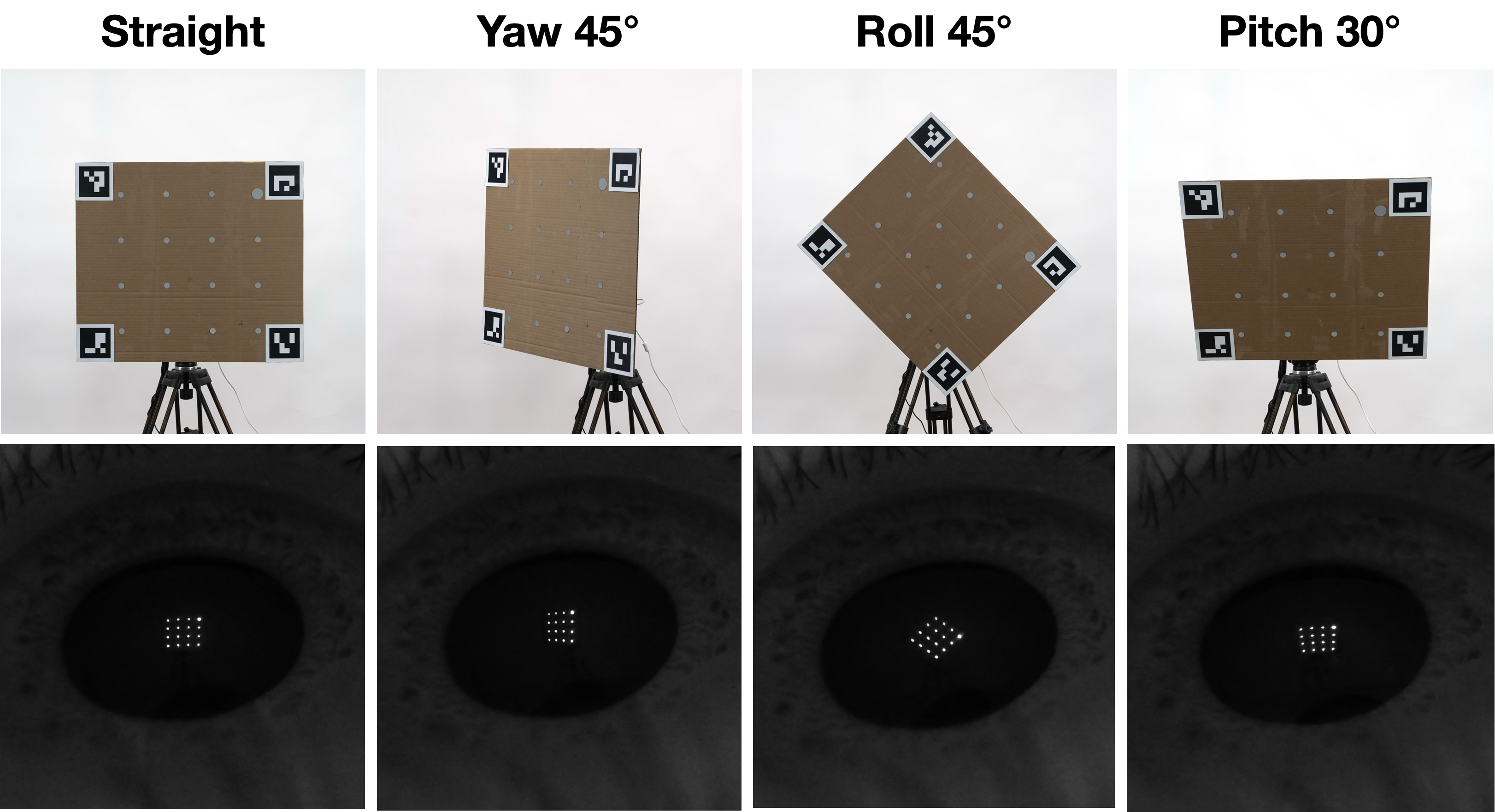}
    \caption{Changes in marker orientation produce corresponding changes in the corneal glint pattern, providing geometric cues for pose estimation. The evaluation board combines a $4\times4$ {\sysname} layout with ArUco markers observed by a separate world-facing camera to obtain reference poses.}
    \label{fig:pnp_example}
\end{figure*}

\subsection{User Calibration}
\label{sec:calibration}

We use calibration eye images with corresponding ground-truth marker orientations and distances to correct errors caused by differences in eye geometry and eyewear fit.

\noindent\textit{Gaze.}
While the user looks at a known marker, we project its center into the eye image using the estimated pose. The difference between this position and the pupil center gives the offset used to correct subsequent gaze coordinates.

\noindent\textit{Orientation.}
We align estimated rotations with ground-truth orientations using a user-specific rotation correction. Viewing the curved cornea from below can distort the glint pattern and bias pitch estimates. We therefore correct the aligned pitch $\widetilde{\theta}$ as
\begin{equation}
    \widehat{\theta}=\alpha\,\widetilde{\theta}+b^{\theta}_u.
    \label{eq:calib_pitch}
\end{equation}
The scale factor $\alpha$ is fitted across users. The offset $b^{\theta}_u$ is the median difference between ground-truth pitch and $\alpha\widetilde{\theta}$ in user $u$'s calibration images.

\noindent\textit{Distance.}
We use a shared calibration curve $f$ to correct nonlinear error in the raw PnP distance. A user-specific scale $s_u$ and offset $b^{d}_u$ adjust for differences in eye geometry and eyewear fit:
\begin{equation}
    \widehat{d}=f\!\left(\frac{\|\mathbf{t}\|}{s_u}\right)+b^{d}_u.
    \label{eq:calib_distance}
\end{equation}
The shared nonlinear correction $f$ combines an isotonic regression fit with a weighted contribution from the scaled input. We fit it using scaled estimates and ground-truth distances across users. The user's scale and offset are estimated from their calibration images and corresponding ground-truth distances.

\section{Evaluation}
We conducted three studies to evaluate {\sysname}. Following human-subjects approval, study~1 examined orientation and distance estimation with 10 participants. The same 10 participants completed Study~2 on object identification. For Study~3, we recruited 6 additional participants following human-subjects approval to evaluate orientation and distance estimation for peripheral {\sysname}. All three studies were conducted under normal indoor lighting.

\subsection{Study 1: Orientation and Distance Estimation}
\label{section:study_pnp_accuracy}

\subsubsection{Study Design}
We constructed an evaluation board with a 16-dot {\sysname} layout ($4 \times 4$ grid, 12~cm patch spacing) and a $2 \times 2$ array of ArUco fiducial markers~(Fig.~\ref{fig:pnp_example}). A separate world-facing camera observed the ArUco markers to provide ground-truth pose via PnP. Using the same setup as the marker characterization study (Fig.~\ref{fig:patches}), a robot arm swept the board through pitch within $\pm30^{\circ}$ and yaw and roll within $\pm45^{\circ}$, at approximately $20^{\circ}$/s.

Each participant completed six sessions: three distances (2, 3, and 4~m), each with the head facing the board (\emph{head aligned}) and turned 20\textdegree{} away (\emph{head offset}). Participants fixated on the marker center in both conditions; only head direction changed. Each session lasted 60~seconds, yielding approximately 900 frames.

To study marker size, we evaluate four-corner subsets of the 16-dot grid spanning 36, 24, and 12~cm, simulating different physical footprints. For each subset, we solve planar PnP using only these geometric anchors and compare against the ArUco ground truth. We report orientation mean absolute error (MAE) in degrees and distance MAE in meters, averaged per user and then across users, with error bars showing the sample standard deviation across participants.

We calibrated orientation using the first 10~s of the 2~m head-aligned recording and distance using the first 5~s at each distance (Sec.~\ref{sec:calibration}). Calibration frames were excluded from scoring; for orientation, we also excluded the next 5~s.

\subsubsection{Orientation Accuracy}

\begin{figure}[t]
    \centering
    \includegraphics[width=0.9\linewidth]{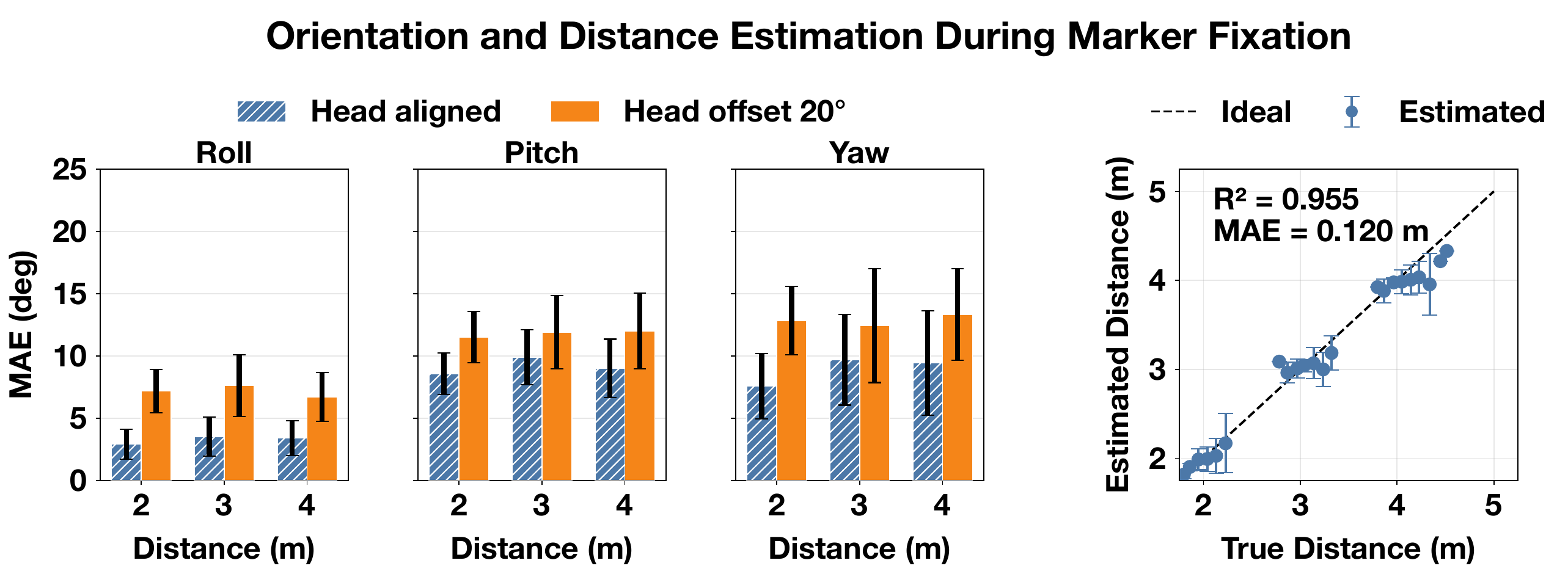}
    \caption{Orientation and distance estimation accuracy. (Left)~Orientation mean absolute error (MAE) for the 24~cm {\sysname} layout with the head aligned or offset by 20\textdegree{}. (Right)~Estimated versus ground-truth distance, combining both head conditions. Error bars show the standard deviation across participants; the dashed line indicates perfect agreement.}
    \label{fig:eval_pnp_distance}
\end{figure}

\noindent \textit{Head Aligned.}
With the 24\,cm {\sysname} layout (Fig.~\ref{fig:eval_pnp_distance}, left), roll is consistently the most accurate axis, with MAE around 3\textdegree{} at all distances. Yaw degrades moderately with distance, from 7.6\textdegree{} at 2\,m to 9.7\textdegree{} at 3\,m and 9.4\textdegree{} at 4\,m. Pitch MAE is around 9--10\textdegree{} across distances.

\noindent \textit{Head Offset 20\textdegree.}
When the head is turned 20\textdegree{} while gaze remains on the marker, all axes show elevated error.
For the 24\,cm {\sysname} layout, roll increases to approximately 7\textdegree, pitch rises to about 12\textdegree, and yaw reaches 12.8\textdegree{} at 2\,m and 13.3\textdegree{} at 4\,m.
Cross-user variability is also wider with the head offset, which may reflect differences in eye geometry and eyewear fit when head and gaze direction diverge.

\subsubsection{Distance Accuracy}
Fig.~\ref{fig:eval_pnp_distance} (right) shows the calibrated distance estimate versus the ArUco ground-truth distance for the 24~cm {\sysname} subset, combining the head-aligned and 20\textdegree{} head-offset conditions. For visualization, we grouped frames into 10~cm bins of ground-truth distance. Each point shows the ground-truth and estimated distances averaged first within each participant and then across participants. Error bars show the standard deviation across participant means. Across 2--4~m, distance estimation achieved an MAE of 0.120~m and an $R^2$ of 0.955.

\subsubsection{Effect of Marker Size}
Table~\ref{tab:marker_size} summarizes accuracy for the three four-corner subsets. Larger layouts reduced orientation error, particularly pitch and yaw at longer distances. At 4~m, increasing the marker span from 12 to 36~cm reduced yaw MAE from 12.28\textdegree{} to 8.91\textdegree{}. Calibrated distance MAE varied by only 0.008~m across sizes. These results suggest that smaller layouts can reduce the required marker footprint with a modest increase in orientation error and little change in distance accuracy.

\begin{table}[t]
    \centering
    \captionsetup{position=bottom,skip=6pt}
    \small
    \setlength{\tabcolsep}{5pt}
    \begin{tabular}{@{}crrrr@{}}
        \toprule
        \textbf{Span (cm)} & \textbf{Roll (\textdegree)} & \textbf{Pitch (\textdegree)} & \textbf{Yaw (\textdegree)} & \textbf{Distance (m)} \\
        \midrule
        12 & 3.64 & 10.23 & 10.33 & 0.124 \\
        24 & 3.28 &  9.16 &  8.90 & 0.120 \\
        36 & 3.01 &  8.90 &  8.25 & 0.116 \\
        \bottomrule
    \end{tabular}
    
    \caption{Mean absolute error (MAE) for four-corner subsets of the evaluation board, averaged over 2--4~m. Orientation results use the head-aligned condition; calibrated distance results combine both head conditions.}
    \label{tab:marker_size}
\end{table}

\subsection{Study 2: Object Identification}
\label{section:study_object_identification}

\begin{figure}[t]
    \centering
    \includegraphics[width=0.7\linewidth]{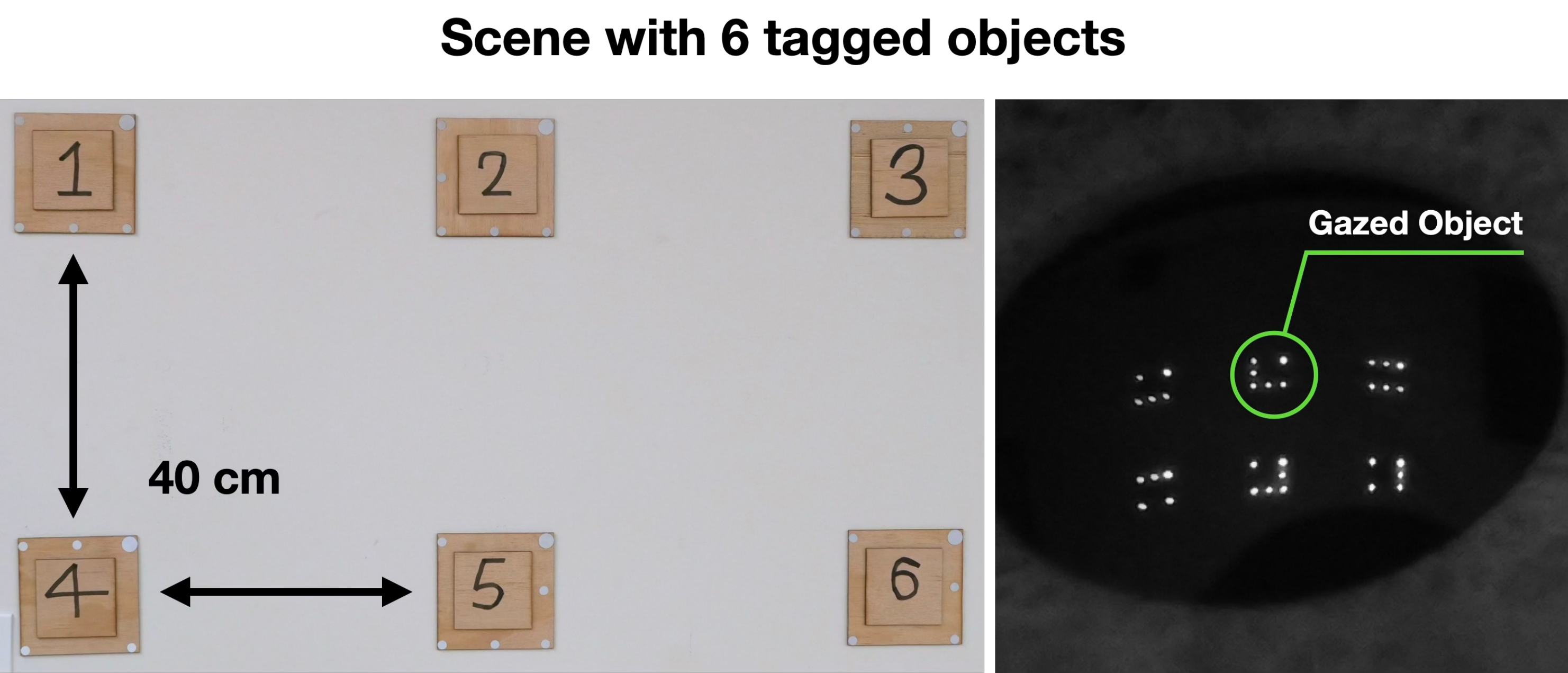}
    \caption{Object-identification study setup with six 15~cm {\sysname} in a $2\times3$ grid (left) and their corneal reflections, with the gazed object highlighted (right).}
    \label{fig:od_example}
\end{figure}

\subsubsection{Study Design}
Six distinct 15~cm {\sysname} were placed 2~m in front of the participant in a $2\times3$ grid, with a 40~cm gap between markers~(Fig.~\ref{fig:od_example}). Each marker used patches 12~mm in diameter and a larger 20~mm anchor patch. Each marker encodes a unique identity using a different retroreflective patch configuration on a shared $3 \times 3$ grid: three markers use 5 patches and three use 6 patches, assigned digits 1 through 6. Identification used a library containing all 32 geometrically distinct patterns achievable on the $3 \times 3$ grid.

Each experimental session consisted of 3 complete sweeps through all 6 markers, yielding 18 trials per participant. In each trial, the system announced a target digit via text-to-speech, and the participant fixated on the corresponding marker for 5~seconds while the eye camera captured frames. The 5~s fixation served only to collect enough frames per condition; the pipeline runs per frame at 15~fps and does not require prolonged fixation. The trial sequence within each sweep was randomized. Only frames captured 3~seconds after digit onset (allowing time for the participant to shift gaze) were included in the analysis. A frame was classified as correct if the predicted gazed-object ID matched the spoken ground-truth digit.

\subsubsection{Results}
\begin{figure}[t]
    \centering
    \includegraphics[width=0.7\linewidth]{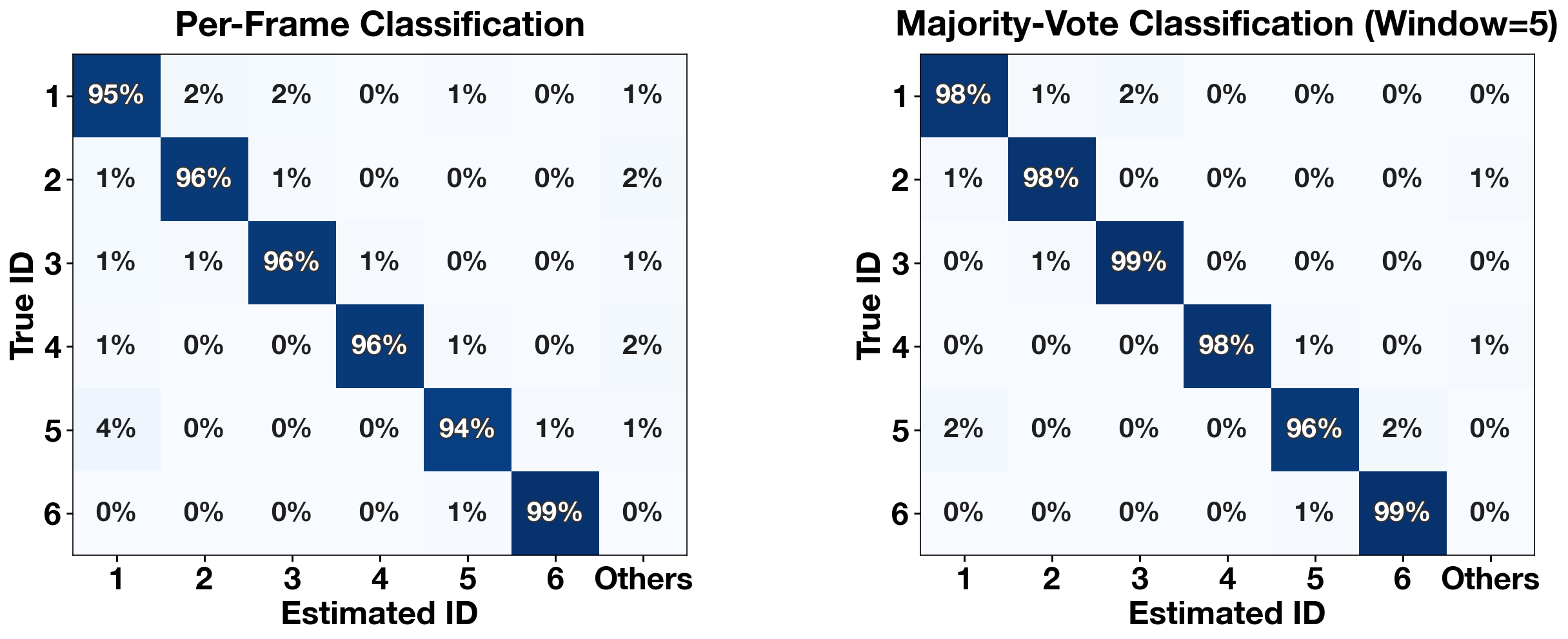}
    \caption{Object identification accuracy per frame (left) and with majority voting over five-frame windows (right). ``Others'' denotes predictions matching one of the 26 non-target library entries.}
    \label{fig:obj_id}
\end{figure}

Fig.~\ref{fig:obj_id} shows the distribution of predicted gazed-object IDs for each target, pooled across participants. The ``Others'' column groups predictions for the 26 non-target IDs in the library.

On a per-frame basis (Fig.~\ref{fig:obj_id}, left), {\sysname} achieves 94--99\% accuracy across all six markers (diagonal entries). Majority voting over non-overlapping windows of 5 consecutive frames (Fig.~\ref{fig:obj_id}, right) improves accuracy to 96--99\% across all markers, with ties broken by the lowest mean PnP reprojection error. Each prediction requires collecting a five-frame window (approximately 0.33~seconds), adding a short delay compared with per-frame prediction.

\subsection{Study 3: Peripheral Markers}
\label{section:study_pnp_ablation}

\subsubsection{Study Design}
We recruited 6 additional participants for this study and used the same board and robot-arm motion as Study~1. Each completed sessions at 2, 3, and 4~m while looking 10\textdegree{} and 20\textdegree{} away from the marker, so the marker was peripheral rather than fixated. Personal calibration uses the same prefix protocol as Study~1, taken from the 10\textdegree{} recordings.

\subsubsection{Results}

\begin{figure}[t]
    \centering
    \includegraphics[width=0.9\linewidth]{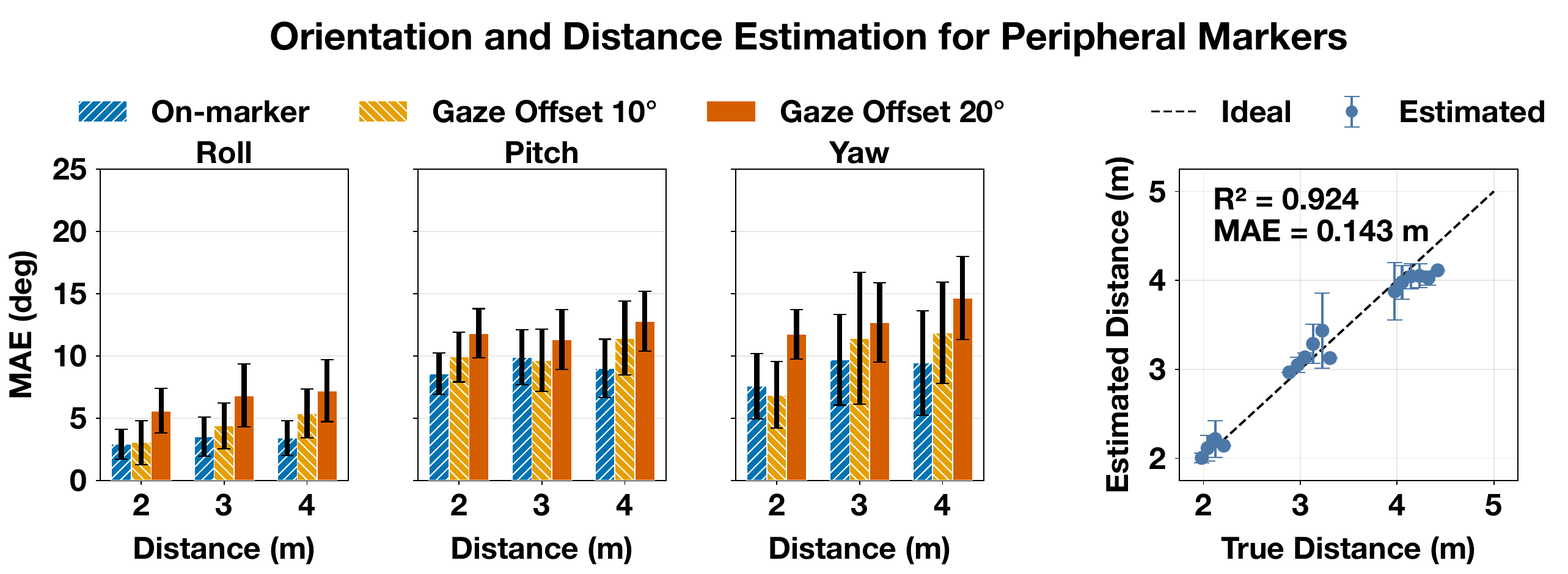}
    \caption{Orientation and distance estimation accuracy for peripheral markers. (Left)~Orientation mean absolute error (MAE) at 10\textdegree{} and 20\textdegree{} gaze offsets (Study~3, $n=6$). Results for direct fixation in Study~1 ($n=10$) provide a reference from a different participant group. (Right)~Estimated versus ground-truth distance, combining both peripheral conditions. Error bars show the standard deviation across participants; the dashed line indicates perfect agreement.}
    \label{fig:eval_pnp_distance_ablation}
\end{figure}

Fig.~\ref{fig:eval_pnp_distance_ablation} (left) shows orientation MAE for the 24\,cm {\sysname} layout at 10\textdegree{} and 20\textdegree{} peripheral viewing, with results from direct fixation in Study~1 included as a reference from different participants. Within Study~3, pitch and yaw errors were higher at 20\textdegree{} than at 10\textdegree{}, while roll remained the most accurate axis. At 20\textdegree{} in the periphery, roll is 5.6--7.2\textdegree, pitch about 12\textdegree, and yaw 11.7\textdegree{} at 2\,m to 14.6\textdegree{} at 4\,m.

Fig.~\ref{fig:eval_pnp_distance_ablation} (right) shows distance with the marker in the periphery, pooling 10\textdegree{} and 20\textdegree{}, using the same binning and averaging procedure as Study~1. Distance estimation achieved an MAE of 0.143~m and an $R^2$ of 0.924. {\sysname} therefore continues to recover orientation and distance when the marker is not the object of gaze.

\section{Discussion}
\label{sec:discussion}
\begin{figure*}
    \centering
    \includegraphics[width=1.0\linewidth]{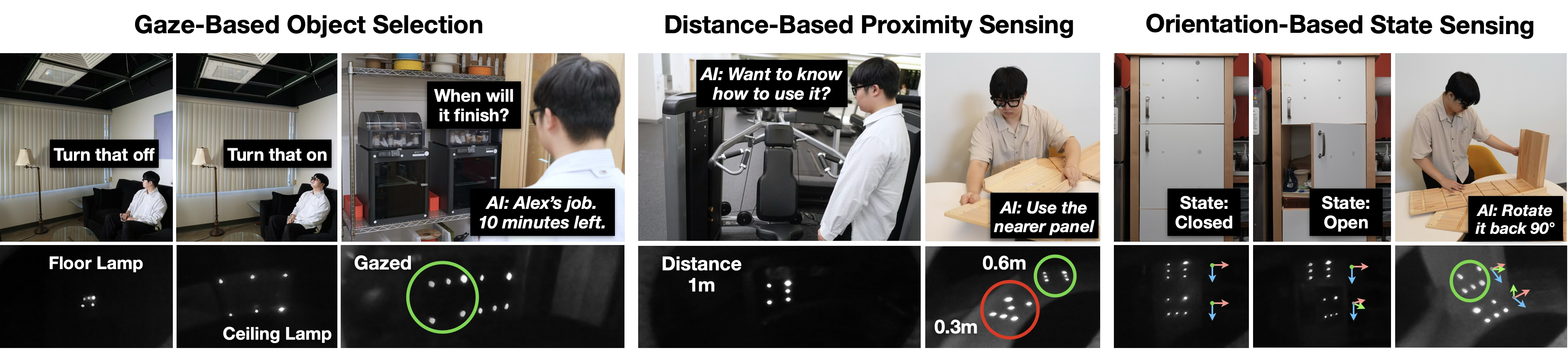}
    \vspace{-2em}
    \caption{Three capabilities of {\sysname} illustrated through application scenarios. Left: gaze and object identity select a lamp or 3D printer for a command or query. Center: distance indicates proximity to equipment and distinguishes nearer from farther parts. Right: orientation supports sensing cabinet-door state and checking the alignment of furniture parts. Callouts show example user requests and assistant responses. The accompanying video demonstrates these application scenarios.}
    \label{fig:application_intent}
\end{figure*}

\subsection{Applications}
\label{sec:applications}

{\sysname} provides context about which object the user is attending to and the distance and orientation of surrounding marked objects. Fig.~\ref{fig:application_intent} illustrates how an assistant could use this context to interpret requests, offer assistance as users approach objects, and provide reminders or guidance based on object state.

\noindent\textit{Gaze-Based Object Selection.}
Combining gaze and speech supports natural references such as ``that,'' reducing the need to name or describe the target~\cite{gazepointar2024}. For example, an assistant could direct a command to the intended lamp or retrieve the job status of a particular printer among visually identical machines, using its marker ID.

\noindent\textit{Distance-Based Proximity Sensing.}
Object identity and distance provide cues about the user's activity, helping an assistant decide when to offer relevant assistance. For example, approaching gym equipment could prompt instructions, while proximity to a coffee maker could prompt a requested reminder about late coffee. During assembly, the assistant could describe the required part as the nearer of two similar panels. Study~3 (Sec.~\ref{section:study_pnp_ablation}) demonstrates distance estimation at 10\textdegree{} and 20\textdegree{} gaze offsets, supporting proximity-based assistance without direct fixation on the marker when its glints remain within the pupil region.

\noindent\textit{Orientation-Based State Sensing.}
Orientation provides cues about an object's physical state and how it changes during use. For example, {\sysname} could detect door or window opening and closing events. Combined with a weather forecast, an observed open window could prompt a reminder before rain. During assembly, known marker placement and a target orientation could guide corrections to misaligned parts. Observed bottle-lid openings could inform supplement reminders, while a kitchen knob's orientation could help an assistant check whether its setting matches the current cooking step and suggest an adjustment.

These applications rely on coarse changes in proximity and object state, for which the measured accuracy could be sufficient. Distance MAE of approximately 12--14~cm over 2--4~m could support approach-based reminders that do not require centimeter-level precision. Everyday events such as doors or windows opening and closing, as well as major part misalignments during assembly, often involve large rotations. The yaw/pitch MAE of 8--15\textdegree{} could therefore be sufficient for detecting such changes. The lower roll MAE of 3--7\textdegree{} could enable finer distinctions, such as tracking changes in kitchen-knob orientation.

\subsection{Privacy}
\label{sec:privacy}

{\sysname} can reduce visual privacy risks in two ways: limiting scene detail during capture and enabling local processing that keeps raw eye images on the glasses.

\noindent\textit{Limiting Scene Capture.}
{\sysname} limits the capture of surrounding scene appearance through NIR filtering and corneal imaging. Eye-tracking cameras use NIR illumination to improve pupil contrast and NIR bandpass filtering to suppress ambient visible light~\cite{kassner2014pupil, lee2015gaze}. In our design, the bandpass filter suppresses visible-light scene reflections before image capture~\cite{midopt_bp850}. The cornea also forms a small, low-contrast image of the surroundings~\cite{nitschke2013corneal}. Retroreflective markers concentrate the glasses' illumination back toward the eye, preserving bright glints against this weak background. Fig.~\ref{fig:application_intent} illustrates how these glints provide object context while much of the detail in the accompanying scene photographs is absent from the eye images. Nearby objects, including those around 1~m, can still appear because ordinary surfaces reflect the glasses' NIR illumination. However, low corneal reflectivity and the small image region occupied by these reflections limit their contrast and spatial detail~\cite{nitschke2013corneal}.

\noindent\textit{Local Processing.}
The computation and power demands of vision models make continuous processing on glasses difficult~\cite{likamwa2014glass, proagent2025}. Some assistants process scene images in the cloud~\cite{raybanmeta_lookandask}, requiring those images to leave the device. After pupil localization, {\sysname} uses lightweight glint detection, marker matching, and geometric pose estimation (Sec.~\ref{sec:software}). Although our prototype uses CNN-based DeepVOG, an efficient pupil detector could replace it~\cite{santini2018pure}. Aria Gen~2 research glasses already estimate pupil position and diameter on-device at up to 90~Hz~\cite{kong2025ariagen2pilot}. Combining efficient pupil localization with our glint-based computations could enable local object sensing, providing only object and gaze information to the assistant while raw eye images are processed and discarded on the glasses.

\subsection{Detection Region}
{\sysname} detects glints within the pupil region, which provides a dark background without iris texture. During Study~3 (Sec.~\ref{section:study_pnp_ablation}), we measured a pupil-reflection field of view of approximately 80\textdegree{} both horizontally and vertically under normal indoor lighting.

Corneal reflections extend beyond the pupil to the region over the iris, which prior work has used to recover scene information~\cite{nishino2006corneal, wang2008separating}. Geometric models indicate that the full cornea can reflect roughly a hemisphere of the surroundings (about 180\textdegree{} across), although camera placement and facial occlusion limit the visible extent~\cite{nishino2006corneal}. Our retroreflective glints are visible over the iris as well, so detecting them throughout the visible cornea could widen the field of view. We leave this extension to future work.

\subsection{Curved Surfaces} 
\begin{figure}
    \centering
    \includegraphics[width=0.4\linewidth]{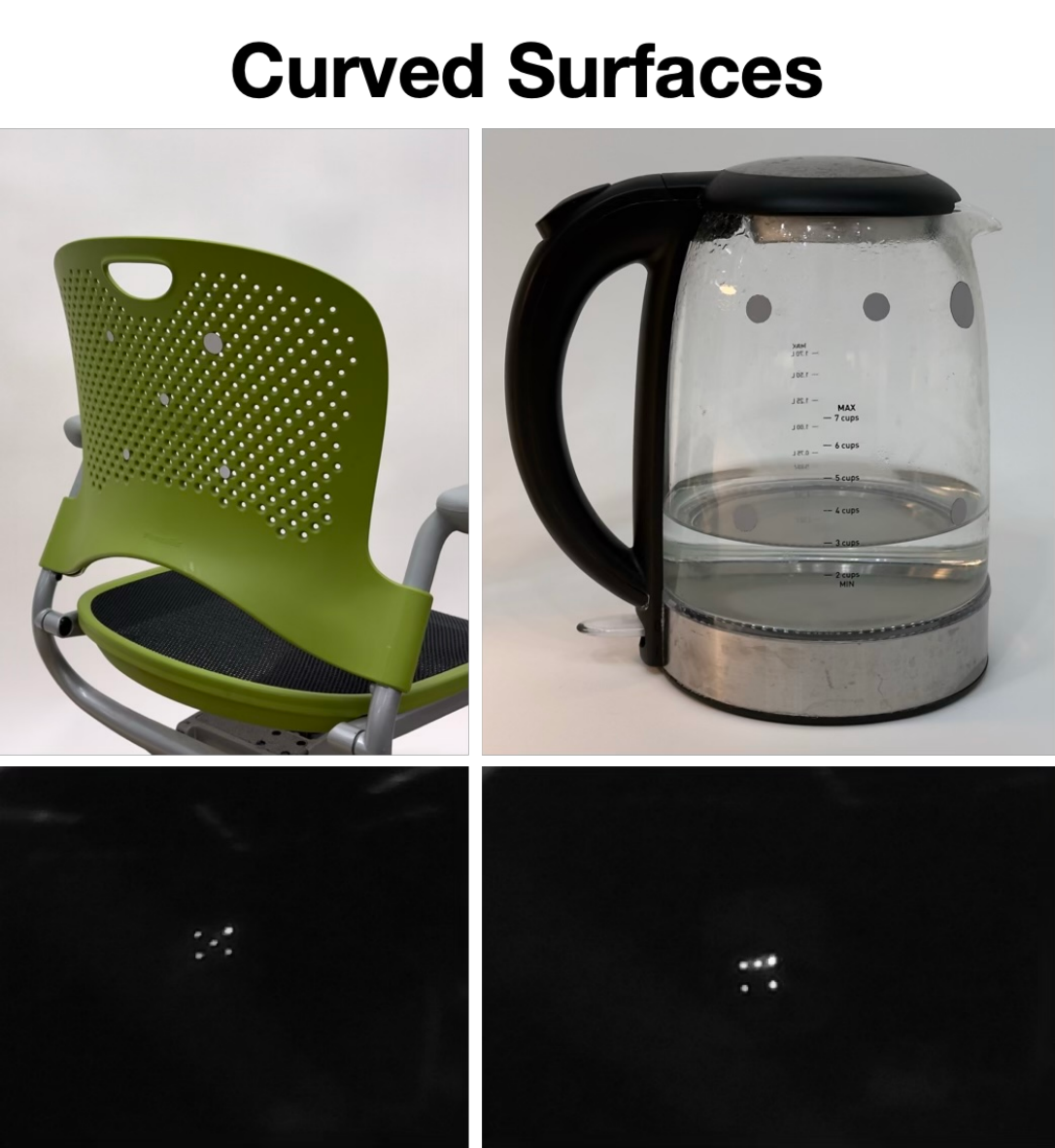}
    \caption{{\sysname} attached to curved objects. Top: examples on a chair and an electric kettle. Bottom: corresponding corneal reflections showing that the markers remain visible on moderately curved surfaces.}
    \label{fig:curved_surfaces}
\end{figure}

We placed {\sysname} on a curved chair back and an electric kettle, viewed directly from 3~m and 2~m, respectively. Both marker patterns remained visible in corneal reflections on these moderately curved surfaces (Fig.~\ref{fig:curved_surfaces}). Stronger curvature could place patches at unfavorable illumination and viewing angles, reducing the light returned toward the glasses.

\subsection{Concealing Markers}
NIR-transmissive housings could allow {\sysname} to be embedded within objects. InfraredTags~\cite{10.1145/3491102.3501951} uses 3D-printing filament that hides embedded patterns in visible light while allowing them to be read by NIR cameras. A similar housing could cover our patches while transmitting both the glasses' illumination and the returning retroreflections. Future work could examine how material choice and housing thickness affect glint brightness and detection range.

\section{Conclusion}
\label{sec:conclusion}

We presented {\sysname}, a system that combines passive retroreflective markers with a single inward-facing eye camera to recover object identity, 3-DoF orientation, distance, and the gazed object. Near-infrared illumination makes the markers visible as corneal glints, and a calibrated pipeline interprets their layout and geometry. Three studies demonstrate object identification and spatial estimation at meter-scale distances, including orientation and distance sensing away from the point of gaze. {\sysname} links gaze, object identity, and spatial context, which could support control, contextual queries, and guidance without a world-facing camera.

\bibliographystyle{ACM-Reference-Format}
\bibliography{sample-base}

\end{document}